\documentstyle[sprocl]{article}
\def\Journal#1#2#3#4{{#1} {\bf #2}, #3 (#4)}
\def\NCA{\em Nuovo Cimento}

\def\NCA{\em Nuovo Cimento}
\def\CR{\em C.R. Acad. Sci. (Paris)}

\def\PRL{\em Phys. Rev. Lett.}
\def\JMP{\em J. Math. Phys.}
\def\GRG{\em Gen. Rel. Grav.}
\def\CQG{\em Class. Quantum Grav.}

\newcommand{\bm}[1]{\mbox{\boldmath $#1$}}

\def\be{\begin{equation}}
\def\ee{\end{equation}}
\def\bea{\begin{eqnarray}}
\def\eea{\end{eqnarray}}
\def\b*{\begin{eqnarray*}}
\def\e*{\end{eqnarray*}}

\def\S{\Sigma}
\def\U{\Upsilon}
\def\R{{\rm I\!R}}
\def\O{\Omega}

\def\DP{{\cal DP}}
\def\SE{{\cal SE}}
\def\SS{{\cal SS}}
\def\NS{{\cal NS}}

\newtheorem{defi}{Definition}[section]
\newtheorem{theo}{Theorem}[section]
\newtheorem{coro}{Corollary}[section]
\newtheorem{prop}{Proposition}[section]
\newtheorem{lem}{Lemma}[section]

\begin{document}

\title{GENERAL ELECTRIC-MAGNETIC DECOMPOSITION OF FIELDS,
          POSITIVITY AND RAINICH-LIKE CONDITIONS}
\author{ JOS\'{E} M.M. SENOVILLA}
\address{Departamento de F\'{\i}sica Te\'orica, Universidad del
Pa\'{\i}s Vasco\\
Apartado 644, 48080 Bilbao, SPAIN\\E-mail: wtpmasej@lg.ehu.es}

\maketitle\abstracts{We show how to generalize the classical electric-magnetic
decomposition of the Maxwell or the Weyl tensors to arbitrary fields described
by tensors of any rank in general $n$-dimensional spacetimes of Lorentzian
signature. The properties and applications of this decomposition are
reviewed. In particular, the definition of tensors quadratic in the original
fields and with important positivity properties is given. These tensors are
usually called ``super-energy" (s-e) tensors, they include the traditional
energy-momentum, Bel and Bel-Robinson tensors, and satisfy the so-called
Dominant Property, which is a straightforward generalization of the classical
dominant energy condition satisfied by well-behaved energy-momentum tensors.
We prove that, in fact, any tensor satisfying the dominant property can be
decomposed as a finite sum of the s-e tensors. Some remarks about the
conservation laws derivable from s-e tensors, with some explicit examples,
are presented. Finally, we will show how our results can be used to
provide adequate generalizations of the Rainich conditions in general
dimension and for any physical field.}

\section{Introduction}
\label{sec:int}
A great deal of theoretical physics is devoted to the {\it
unification} of concepts. Two rather early and
outstanding examples of physical unifications are given by the
concepts of
\begin{center}
    \fbox{\underline{ELECTRO}-MAGNETISM \hspace{7mm} and \hspace{7mm}
    SPACE-\underline{TIME}}
\end{center}
which respectively unify the electric field with the magnetic field,
and the space and time measurements. Of course, these two
unifications are obviously related, and I shall try to maintain the
coherence by underlying the corresponding concepts (so that
`electric' is related to `time'). Thus, for instance, the
\underline{electro}magnetic field is simply the space\underline{time}
unification of the classical electric field $\vec E$ with the
classical magnetic induction $\vec H$.

Nevertheless, it is very important to keep in mind that we do not
measure ``spacetime intervals'', nor ``electromagnetic strengths''.
Actually, in labs we can only measure
\begin{center}
\begin{table}[htbp]\centering
\begin{tabular}{|l|c|}
\hline  Space distances & \underline{Time} intervals\\
\hline
\underline{Electric} field & Magnetic field \\ \hline
\end{tabular}
\end{table}
\end{center}
by using rods, clocks, electrometers, Gaussmeters or fluxmeters. This
means that we must know how to do the {\it splitting} of the unified
concepts. So we need to define how to do the followign breakings
\begin{center}
    space\underline{time} $\left\{ \begin{tabular}{l}
    \mbox{space} \\
    \mbox{\underline{time}}
    \end{tabular}\right. \, ,
    \hspace{1cm}
    \mbox{\underline{electro}magnetic} \left\{ \begin{tabular}{l}
    \mbox{\underline{electric}} \\ \mbox{magnetic}
    \end{tabular}\right.$
\end{center}
which, naturally, are intimately related and, more importantly, depend
on the observer. Let us remind that an observer is defined, in General
Relativity and similar geometrical theories, by a unit timelike vector
field $\vec u$. In the domain of validity of the observer described
by $\vec u$ one has that
\begin{enumerate}
  \item The space/\underline{time} splitting leads to the theory of
  reference frames. The vector field $\vec u$ defines a congruence of
  timelike curves $\cal C$ so that, for this observer, the
  infinitessimal variations of time $T$ are represented somehow by the
  1-form $u_{\mu}\propto$ ``$\delta_{\mu} T\,$'', while the corresponding
  space is defined by the quotient $V_{n}/\cal C$ of the manifold by
  the congruence. Let us remark that, according to some authors
  \cite{Belcong,Mashhoon}, and in my opinion too, the metric structure of
  this space is debatable, so that a unequivocal way to measure
  distances (without using light rays) is not clearly defined.
  \item The \underline{electric}/magnetic decomposition has not been
  pursued in full generality until recently \cite{S,S2,S3}. This general
  decomposition, for any given tensor, will be presented and briefly
  analized here in any spacetime of {\it arbitrary dimension} $n$.
\end{enumerate}

The only pre-requisite we will need in order to achieve the
\underline{electric}/magnetic decomposition is the existence of a
metric $g$ with Lorentzian signature (i.e., the possibility of defining
time), which we will take to be (--,+,\dots ,+). Apart from that, our
considerations are completey general and therefore applicable to
almost any available geometric physical theory.

\subsection{The \underline{electro}magnetic field as guide \dots}
\label{subsec:e-m}
Let us take, for a moment, the typical case of an electromagnetic
field\footnote{I use the standard square and
round brackets to denote the usual (anty-) symmetrization.}
$F_{\mu\nu}=F_{[\mu\nu]}$
in a 4-dimensional Lorentzian manifold $(V_{4},g)$. The
\underline{electric} $\vec E(\vec u)$ and the magnetic $\vec H(\vec u)$ fields
relative to the observer $\vec u$ are defined at any
point by
\b*
E_{\mu}(\vec u)\equiv F_{\mu\nu}u^{\nu}, \hspace{1cm}
H_{\mu}(\vec u)\equiv \stackrel{*}{F}_{\mu\nu}u^{\nu}
\e*
respectively, where $\stackrel{*}{F}_{\mu\nu}\equiv (1/2)
\eta_{\mu\nu\rho\sigma}F^{\rho\sigma}$ is the Hodge dual of
$F_{\mu\nu}$ constructed by using the volume element 4-form
$\bm{\eta}$. Obviously we have that
\b*
E_{\mu}u^{\mu}=H_{\mu}u^{\mu}=0
\e*
so that the electric and magnetic fields are {\it spatial vectors}
relative to the observer $\vec u$. This implies that $\vec E(\vec u)$ and
$\vec H(\vec u)$ have 3 independent components each, which altogether
add up to the 6 independent components of the 2-form $F_{\mu\nu}$
in 4 dimensions. These vectors allow to define the following classes of
electromagnetic fields
\b*
F_{\mu\nu}\,\, \mbox{is called}\, \left\{\begin{tabular}{ccc}
purely electric & if & $\exists \vec u \,\, \mbox{such that}\,\,\vec
H(\vec u)=\vec 0$ \\
purely magnetic & if & $\exists \vec u \,\, \mbox{such that}\,\,\vec
E(\vec u)=\vec 0$ \\
null & if & $F_{\mu\nu}\stackrel{*}{F}{}^{\nu}{}_{\rho}=0\, \,
\mbox{and}\,\, F_{\mu\nu}F^{\mu\nu}=0$ \end{tabular}\right.
\e*
Mind, however, that not all electromagnetic fields belong to one of these
three classes. In all three types the condition $E_{\mu}H^{\mu}=0$
holds, as this is an invariant: $\forall \vec u , \,\, 2E_{\mu}H^{\mu}=
F_{\mu\nu}\stackrel{*}{F}{}^{\mu\nu}$. The other invariant
$E_{\mu}E^{\mu}-H_{\mu}H^{\mu}=-F_{\mu\nu}F^{\mu\nu}$ selects the
particular type according to sign.

The energy-momentum tensor of the electromagnetic field in 4
dimensions is defined by
\be
\tau_{\mu\nu}\equiv \frac{1}{2}\left(F_{\mu\rho}F_{\nu}{}^{\rho}+
\stackrel{*}{F}_{\mu\rho}\stackrel{*}{F}_{\nu}{}^{\rho}\right),
\label{emem4}
\ee
and satisfies
\be
\tau_{\mu\nu}=\tau_{(\mu\nu)}, \hspace{5mm} \tau_{\mu}{}^{\mu}=0,
\hspace{5mm} \tau_{\mu\rho}\tau^{\rho}{}_{\nu}\propto g_{\mu\nu}.
\label{raiem}
\ee
Conditions (\ref{raiem}) are the basis of the so-called Rainich 
theory, to be considered in more generality in sections \ref{sec:dp} and 
\ref{sec:rainich}.
By using Maxwell equations, $\nabla_{[\rho}F_{\mu\nu]}=0$,
$\nabla_{\rho}F^{\rho\mu}=-j^{\mu}$ where $\vec j$ is the 4-current,
one derives $\nabla_{\rho}\tau^{\rho\mu}=F^{\mu\nu}j_{\nu}$ from where
\b*
\vec j =\vec 0 \hspace{5mm} \Longrightarrow \hspace{5mm}
\nabla_{\rho}\tau^{\rho\mu}=0.
\e*
If there is a (conformal) Killing vector $\vec \zeta$, then the last
expression leads to
\b*
\nabla_{\rho}J^{\rho}=0, \hspace{1cm} J^{\rho}(F;\zeta)\equiv
\tau^{\rho\mu}\zeta_{\mu}
\e*
so that the {\it divergence-free vector} (also called current)
$\vec J$ can be used to construct conserved quantities via Gauss
theorem.

The energy density of the electromagnetic field with respect to the
observer $\vec u$ is defined by
\be
\Omega (\vec u)\equiv \tau_{\mu\nu}u^{\mu}u^{\nu}=\frac{1}{2}
\left(E_{\mu}E^{\mu}+H_{\mu}H^{\mu}\right) \label{e-me}
\ee
from where we immediately deduce $\Omega (\vec{u})\geq 0$
for all timelike $\vec{u}$ and
\b*
\left\{\exists \vec{u}\hspace{3mm} \mbox{such that} \hspace{2mm}
\Omega\left(\vec{u}\right)=0 \right\} \Longleftrightarrow
\tau_{\mu\nu}=0 \Longleftrightarrow F_{\mu\nu}=0.
\e*
Actually, $\tau_{\mu\nu}$ satisfies a stronger condition, usually called
the dominant energy condition,\cite{Ple,HE} given by
\be
\tau_{\mu\nu}v^{\mu}w^{\nu}\geq 0 \hspace{5mm} \mbox{for all causal
future-pointing}\,\,\, \vec{v}, \,\, \vec{w} \label{dec}
\ee
which is equivalent to saying that the flux energy vector
$p^{\mu}(\vec u)\equiv -\tau^{\mu}{}_{\nu}u^{\nu}$ is always causal and future
pointing for all possible observers.

\subsection{\dots and the curvature tensors as inspiration.}
\label{subsec:weyl}
In General Relativity, there is another well-known `E-H'
decomposition, namely, that of the Weyl tensor, introduced in
particular coordinates in \cite{Mat} and then in general in
\cite{XYZZ}, see also \cite{Beltesis,Old}. In $n$ dimensions,
the Weyl tensor reads
\b*
C_{\alpha\beta,\lambda\mu}=R_{\alpha\beta,\lambda\mu}-\frac{2}{n-2}\left(
R_{\alpha[\lambda}g_{\mu]\beta}-R_{\beta[\lambda}g_{\mu]\alpha}\right)+
\frac{2\, R}{(n-1)(n-2)}g_{\alpha[\lambda}g_{\mu]\beta}
\e*
where $R_{\alpha\beta\lambda\mu}$ is the Riemann tensor,
$R_{\alpha\lambda}\equiv R^{\rho}{}_{\alpha\rho\lambda}$ is the Ricci
tensor and $R\equiv R\equiv R^{\rho}{}_{\rho}$ is the scalar
curvature, and it has the same
symmetry properties as the Riemann tensor but is also traceless
\b*
C_{\alpha\beta\lambda\mu}=C_{[\alpha\beta][\lambda\mu]}, \hspace{3mm}
C_{\alpha[\beta\lambda\mu]}=0, \hspace{3mm} C^{\rho}{}_{\beta\rho\mu}=0.
\e*
Due to the Lanczos identity, see e.g.\cite{S3}, the Weyl tensor has
only one independent Hodge dual \underline{in 4 dimensions}, given by
$\stackrel{*}{C}_{\alpha\beta\mu\nu}\equiv (1/2)\eta_{\mu\nu\rho\sigma}
C_{\alpha\beta}{}^{\rho\sigma}=(1/2)\eta_{\alpha\beta\rho\sigma}
C^{\rho\sigma}{}_{\mu\nu}$. Therefore, one can define the `electric'
and `magnetic' parts of the Weyl tensor with regard to $\vec u$
respectively as (in $n=4$ !)
\bea
E_{\alpha\lambda}(\vec u)\equiv
C_{\alpha\beta\lambda\mu}u^{\beta}u^{\mu}, \hspace{4mm}
E_{\alpha\lambda}=E_{\lambda\alpha}, \hspace{4mm} E^{\mu}{}_{\mu}=0,
\hspace{4mm} u^{\alpha}E_{\alpha\lambda}=0, \label{WE}\\
H_{\alpha\lambda}(\vec u)\equiv
\stackrel{*}{C}_{\alpha\beta\lambda\mu}u^{\beta}u^{\mu}, \hspace{4mm}
H_{\alpha\lambda}=H_{\lambda\alpha}, \hspace{4mm} H^{\mu}{}_{\mu}=0,
\hspace{4mm} u^{\alpha}H_{\alpha\lambda}=0, \label{WH}
\eea
so that they are spatial symmetric traceless tensors. Each of them has
5 independent components adding up to the 10 components of
$C_{\alpha\beta\lambda\mu}$, see\cite{S3,Old,MB} for more details. Less
known but equally old is the similar decomposition of the
full Riemann tensor, also introduced in\cite{XYZZ}, and given by the
following {\it four} 2-index tensors (in $n=4$ !)
\b*
&Y_{\alpha\lambda}\left(\vec{u}\right)\equiv R_{\alpha\beta\lambda\mu}
u^{\beta}u^{\mu}, \hspace{6mm}
X_{\alpha\lambda}\left(\vec{u}\right)\equiv {*R*}_{\alpha\beta\lambda\mu}
u^{\beta}u^{\mu}, \\
&Z_{\alpha\lambda}\left(\vec{u}\right)\equiv {*R}_{\alpha\beta\lambda\mu}
u^{\beta}u^{\mu},\hspace{6mm} Z'_{\alpha\lambda}\left(\vec{u}\right)\equiv
{R*}_{\alpha\beta\lambda\mu}  u^{\beta}u^{\mu}
\e*
where the place of the * indicates the pair of skew indices which are
Hodge-dualized, see\cite{XYZZ,Beltesis,BS2}. Taking into
account the symmetry properties of $R_{\alpha\beta\lambda\mu}$,
they satisfy\cite{XYZZ,Beltesis,BS2}
\begin{eqnarray*}
&X_{\alpha\lambda}=X_{\lambda\alpha},\hspace{6mm}
u^{\alpha}X_{\alpha\lambda}=0,\hspace{6mm}
Y_{\alpha\lambda}=Y_{\lambda\alpha},\hspace{6mm}
u^{\alpha}Y_{\alpha\lambda}=0,\hspace{6mm} \\
&Z_{\alpha\lambda}=Z'_{\lambda\alpha}, \hspace{6mm}
u^{\alpha}Z_{\alpha\lambda}=0,\hspace{6mm}
u^{\alpha}Z'_{\alpha\lambda}=0,\hspace{6mm}
Z^{\alpha}_{\alpha}=Z'^{\alpha}_{\alpha}=0
\end{eqnarray*}
so that $X_{\alpha\lambda}$ and $Y_{\alpha\lambda}$ have 6 independent
components each, while $Z_{\alpha\lambda}$ (or
equivalently $Z'_{\alpha\lambda}$) has the remaining 8 independent
components of the Riemann tensor. They wholly determine the Riemann tensor.

It is important to stress here that the above properties and
decompositions depend {\it crucially on the dimension} of the
spacetime, and many of the above simple things are no longer true in
general dimension $n$. As perhaps a surprising example let us remark
that the `E-H' decomposition of the Weyl tensor has in general
{\it four} different spatial tensors (as that of the Riemann tensor
above). All this will be analyzed in the next section and clarified in
subsection \ref{examples}.

\section{Tensors as $r$-folded forms: general E-H decompositions.}
\label{sec:EH}
In order to define a fully general `electric-magnetic' decomposition of
any $m$-covariant tensor $t_{\mu_1\dots\mu_m}$, the key idea is to
split its indices into antisymmetric blocks (say $r$ in total)
denoted generically by $[n_{\U}]$, where $n_{\U}$ is the
number of antisymmetric indices in the block and $\U=1,\dots,r$,
see\cite{S,S3}. In this way, we look at $t_{\mu_1\dots\mu_m}$
as an {\it r-fold $(n_1,\dots,n_r)$-form},
where obviously $n_1+\dots +n_r =m$, and use the notation
$t_{[n_1],\dots,[n_r]}$ for the tensor.

Then, one constructs all the duals by using the Hodge *
with the volument element $\eta_{\mu_1\dots\mu_n}$ acting on each
of the $[n_{\U}]$ blocks, obtaining a total of $2^r$ different
tensors, each of which is
an $r$-fold form (except when $n_{\U}=n$ for some $\U$).
I define the
canonical `E-H' decomposition of $t_{\mu_1\dots\mu_m}$
relative to $\vec u$ by contracting each of these duals on all
their $r$ blocks with $\vec{u}$\cite{S,S3}: whenever $\vec{u}$ is contracted
with a `starred' block $\stackrel{*}{[n-n_{\U}]}$, we get a `magnetic part'
in that block, and an `electric part' otherwise. Thus, the electric-magnetic
parts are (generically) $r$-fold forms which can be denoted by
\b*
({}^t_{\vec{u}}\underbrace{EE\dots E}_r)_{[n_1-1],[n_2-1],\dots,[n_r-1]},\,\,
({}^t_{\vec{u}}H\underbrace{E\dots
E}_{r-1})_{[n-n_1-1],[n_2-1],\dots,[n_r-1]},\,\,
\dots \\
({}^t_{\vec{u}}\underbrace{E\dots
E}_{r-1}H)_{[n_1],\dots,[n_{r-1}-1][n-n_r-1]},\,
({}^t_{\vec{u}}HH\underbrace{E\dots
E}_{r-2})_{[n-n_1-1],[n-n_2-1],\dots,[n_r-1]},
\\
\dots ,
({}^t_{\vec{u}}\underbrace{E\dots
E}_{r-2}HH)_{[n_1-1],\dots,[n-n_{r-1}-1],[n-n_r-1]}\, ,
\, \dots \hspace{15mm}\\
\dots ,\dots ,
({}^t_{\vec{u}}\underbrace{HH\dots H}_r)_{[n-n_1-1],[n-n_2-1],\dots,[n-n_r-1]},
\e*
where, for instance,
\b*
({}^t_{\vec{u}}\underbrace{EE\dots E}_r)_{\mu_2\dots\mu_{n_1},
\dots ,\rho_2\dots\rho_{n_r}}\equiv
\tilde{t}_{\mu_1\mu_2\dots\mu_{n_1},
\dots ,\rho_1\rho_2\dots\rho_{n_r}}u^{\mu_1}\dots u^{\rho_1}\, ,\\
({}^t_{\vec{u}}H\underbrace{E\dots E}_{r-1})_{\mu_{n_1+2}\dots\mu_{n},
\dots ,\rho_2\dots\rho_{n_r}}\!\equiv\!
\tilde{t}_{\stackrel{*}{\mu_{n_1+1}\mu_{n_1+2}\dots\mu_{n}},
\dots ,\rho_1\rho_2\dots\rho_{n_r}}u^{\mu_{n_1+1}}\dots u^{\rho_1} = \\
=\eta_{\mu_1\mu_2\dots\mu_n}\tilde{t}^{\mu_1\mu_2\dots\mu_{n_1},}
{}_{\dots ,\rho_1\rho_2\dots\rho_{n_r}}u^{\mu_{n_1+1}}\dots u^{\rho_1}
\hspace{15mm}
\e*
and so on. Here, $\tilde{t}_{[n_1],\dots,[n_r]}$ denotes the tensor obtained
from $t_{[n_1],\dots,[n_r]}$ by permutting the indices such that
they are in the order given by $[n_1]\dots [n_r]$. In general,
there are $2^r$ independent E-H parts, they are {\it spatial} relative to
$\vec{u}$ in the sense that they are orthogonal to $\vec{u}$ in any index,
and all of them determine $t_{\mu_1\dots\mu_m}$ completely.
Besides, $t_{\mu_1\dots\mu_m}$ vanishes iff
all its E-H parts do.

A special case of relevance is that of {\it decomposable} $r$-fold 
forms: $t_{[n_1],\dots,[n_r]}$ is said to be decomposable
if there are $r$ forms $\O^{(\U)}_{\mu_1\dots \mu_{n_{\U}}}=
\O^{(\U)}_{[\mu_1\dots \mu_{n_{\U}}]}$ ($\U=1\dots r$)
such that
\be
\tilde{t}_{[n_1] \dots [n_r]}=\left(\O^{(1)}\otimes \dots \otimes
\O^{(r)}\right)_{[n_1] \dots [n_r]}.\label{decom}
\ee
Obviously, any $r$-fold form is a sum of decomposable ones.

\subsection{Single blocks or single $p$-forms}
\label{subsec:singleEH}
In order to see what the `E-H' decomposition accomplishes, let us
concentrate on a single block $[n_\U]$ with $n_{\U}=p$ indices or,
for the sake of clarity and simplicity, on a single $p$-form
$\S_{\mu_1\dots\mu_p}=\S_{[\mu_1\dots\mu_p]}$. In general, $\bm{\S}$
has $\left(\begin{array}{c} n\\ p\end{array}\right)$ independent components.
The electric part of $\bm{\S}$ with respect
to $\vec{u}$ is the $(p-1)$-form
\b*
\left({}^{\S}_{\vec{u}}E\right)_{\mu_2\dots\mu_p}=
\left({}^{\S}_{\vec{u}}E\right)_{[\mu_2\dots\mu_p]}\equiv
\S_{\mu_1\mu_2\dots\mu_p}u^{\mu_1}, \hspace{3mm}
\left({}^{\S}_{\vec{u}}E\right)_{\mu_2\dots\mu_p}u^{\mu_2}=0.
\e*
Obviously, $\left({}^{\S}_{\vec{u}}\bm{E}\right)$ has
$\left(\begin{array}{c} n-1 \\ p-1 \end{array}\right)$ independent components,
which correspond to those with a `zero' among the components
$\S_{\mu_1\dots\mu_p}$ in any orthonormal basis
$\{\vec{e}_{\mu}\}$ with $\vec{e}_0=\vec{u}$, that is, to $\S_{0i_2\dots i_p}$
(we use italic lower-case indices $i,j,\dots =1,\dots, n-1$). Similarly,
the magnetic part of $\bm{\S}$ with respect to $\vec{u}$ is the $(n-p-1)$-form
\b*
\left({}^{\S}_{\vec{u}}H\right)_{\mu_{2}\dots\mu_{n-p}}=
\left({}^{\S}_{\vec{u}}H\right)_{[\mu_{2}\dots\mu_{n-p}]}\equiv
\S_{\stackrel{*}{\mu_{1}\mu_{2}\dots\mu_{n-p}}}u^{\mu_{1}},\,\,
\left({}^{\S}_{\vec{u}}H\right)_{\mu_{2}\dots\mu_{n-p}}u^{\mu_2}=0.
\e*
$\left({}^{\S}_{\vec{u}}\bm{H}\right)$ has
$\left(\begin{array}{c} n-1 \\ n-p-1 \end{array}\right)=
\left(\begin{array}{c} n-1 \\ p \end{array}\right)$ independent components,
which correspond to those without `zeros' in the basis above, that is,
to $\S_{i_1\dots i_{p}}$. The sum of the components of the electric
part plus the components of the magnetic part gives
\b*
\left(\begin{array}{c} n-1 \\ p-1 \end{array}\right)+
\left(\begin{array}{c} n-1 \\ p \end{array}\right)=
\left(\begin{array}{c} n\\ p\end{array}\right)
\e*
which is the total number of components of $\bm{\S}$. The single
$p$-form $\bm{\S}$ can be expressed in terms
of its electric and magnetic parts as\cite{S3}
\b*
\S_{\mu_1\dots\mu_p}=-p\, u_{[\mu_1}({}^{\S}_{\vec{u}}E)_{\mu_2\dots\mu_p]}+
\frac{(-1)^{p(n-p)}}{(n-p)!}\eta_{\rho_1\dots\rho_{n-p}\mu_1\dots\mu_p}
u^{\rho_1}\left({}^{\S}_{\vec{u}}H\right)^{\rho_{2}\dots\rho_{n-p}}\\
\S_{\stackrel{*}{\mu_{p+1}\mu_{p+2}\dots\mu_n}}=
-(n-p)u_{[\mu_{p+1}}({}^{\S}_{\vec{u}}H)_{\mu_{p+2}\dots\mu_n]}-\frac{1}{p!}
\eta_{\mu_1\dots\mu_n}u^{\mu_1}\left({}^{\S}_{\vec{u}}E
\right)^{\mu_2\dots\mu_p}
\e*
which can be rewritten in a compact form as
\b*
\bm{\S}=-\bm{u}\wedge \left({}^{\S}_{\vec{u}}\bm{E}\right)+(-1)^{p(n-p)}
\stackrel{\hspace{-1mm}*}{\left[\bm{u}\wedge \left({}^{\S}_{\vec{u}}\bm{H}
\right)\right]},
\hspace{5mm}
\stackrel{*}{\bm{\S}}=-\bm{u}\wedge \left({}^{\S}_{\vec{u}}\bm{H}\right)-
\stackrel{*}{\left[\bm{u}\wedge \left({}^{\S}_{\vec{u}}\bm{E}\right)\right]}
\e*
where $\wedge$ denotes the exterior product. These formulae show
that the E-H parts of $\bm{\S}$ determine $\bm{\S}$, and that $\bm{\S}$
vanishes iff its E-H parts are zero with respect to some (and
then to all) $\vec{u}$. The general E-H decomposition for
arbitrary $r$-fold forms does all this on each $[n_{\U}]$-block.

Obviously, one can put forward the following definition: a single
$p$-form (or alternatively a single $[p]$ block of an $r$-fold form)
\b*
\S_{[p]}\,\, \mbox{is called}\, \left\{\begin{tabular}{l}
purely electric if $\exists \vec u \,\, \mbox{such that}\,
\left({}^{\S}_{\vec{u}}\bm{H}\right)=\bm{0}$ \\
purely magnetic if $\exists \vec u \,\, \mbox{such that}\,
\left({}^{\S}_{\vec{u}}\bm{E}\right)=\bm{0}$ \\
null if $\S_{\rho\mu_{2}\dots\mu_{p}}
\stackrel{*}{\S}{}^{\rho}{}_{\nu_{2}\dots\nu_{n-p}}=0\, \,
\mbox{and}\,\, \S_{\mu_{1}\dots\mu_{p}}\S^{\mu_{1}\dots\mu_{p}}=0$
\end{tabular}\right.
\e*

\subsection{Illustrative examples}
\label{examples}
As initial example, take a single
1-form $A_{\mu}$ with components $(A_{0},A_{1},\dots,A_{n-1})$ in any
orthonormal (ON) basis $\{\vec{e}_{\mu}\}$. Its electric part with respect
to $\vec{u}=\vec{e}_0$ is simply $A_{0}$ (that is, the time component),
and the corresponding magnetic part is $(0,A_{1},\dots,A_{n-1})$ (the
spatial part). As is obvious, $\bm{A}$ is purely electric, purely
magnetic or null according to whether the vector $\vec A$ is
timelike, spacelike or null, respectively. If we now consider a
2-form $F_{[2]}$ in the given ON basis
\b*
\left(F_{\mu\nu}\right)=\left(\begin{array}{lcccc}
0 & \underline{F_{01}} & \underline{F_{02}} & \dots & \underline{F_{0\,n-1}} \\
  & 0      & F_{12} & \dots & F_{1\,n-1} \\
  &        & 0      & \dots & F_{2\,n-1} \\
  &        &        & \dots & \dots       \\
  &        &        &       &  0
\end{array}\right)
\e*
its electric part is the spatial 1-form constituted by the underlined
components while the rest of components provide the magnetic part,
which is a spatial $(n-3)$-form.

Going to the case of 2-fold (or double) forms, and starting with the
simplest case of a double (1,1)-form $A_{\mu\nu}$
($A_{(\mu\nu)}\neq 0$), the four E-H parts can be identified by
writing $A_{[1][1]}$ in the ON basis above
\b*
\left(A_{\mu\nu}\right)=\left(\begin{array}{ccccc}
\fbox{$A_{00}$} & \underline{A_{01}} & \underline{A_{02}} & \dots &
\underline{A_{0\,n-1}} \\
\underbrace{A_{10}}  & A_{11}      & A_{12} & \dots & A_{1\,n-1} \\
\underbrace{A_{20}}  & A_{21}       & A_{22}      & \dots & A_{2\,n-1} \\
\dots & \dots    & \dots   & \dots & \dots       \\
\underbrace{A_{n-1\,0}}  & A_{n-1\,1}  & A_{n-1\,2} & \dots & A_{n-1\,n-1}
\end{array}\right)
\e*
so that the `EE' part is given by the scalar $A_{00}$ in the box, the
`EH' part is given by the underlined components, the `HE' part is
constituted by the underbraced components, and the `HH' part is the
double $(n-2,n-2)$-form given by the rest. In the case that
$A_{\mu\nu}=A_{(\mu\nu)}$ is symmetric, the EH and HE parts are equal
and the HH part is symmetric too. Some particular cases are
of interest here. For instance, the metric itself is a double
(1,1)-form and admits its own E-H decomposition. Its EE part is the
`$g_{00}$' component (related to the Newtonian potential), and its
EH=HE part, corresponding to the `$g_{0i}$' in general, can always be
set to zero by choice of orthonormal basis, and as is
known, these ON basis can be further chosen to be {\it locally}
holonomous (Gaussian coordiantes). At this level, this means that the
gravimagnetic effects are always {\it locally} inertial, that is, due
to the reference frame. Another case of interest is the general
energy-momentum tensor $T_{\mu\nu}$. Comparing with its canonical
decomposition along $\vec u$ (see e.g. \cite{MB}), its EE part
corresponds to the energy density, its EH=HE part is the energy-flux
vector, and the HH part is given by the pressures and/or tensions, all
of these with regard to $\vec u$.

Let us consider finally the case of double (2,2)-forms, which in
particular contain the curvature tensors. In general, such a
$K_{[2][2]}$ has $n^2(n-1)^2/4$ independent components,
and they are distributed as follows: its EE part is a double
(1,1)-form carrying $(n-1)^2$ independent components, its EH and HE
parts are double $(1,n-3)$- and $(n-3,1)$-forms respectively and
have $(n-1)^2(n-2)/2$ components each, and its HH part is a double
$(n-3,n-3)$-form with the remaining $(n-1)^2(n-2)^2/4$ components.
However, some particular cases arise for the curvature tensors
because, for instance, the Riemann tensor satisfies the first Bianchi
identity $R_{\alpha[\mu\nu\rho]}=0$. This implies that $R_{[2][2]}$
has in fact $n^2(n^2-1)/12$ independent components, which are
distributed into $n(n-1)/2$ components for the EE part,
$n(n-1)(n-2)/3$ for the EH (which is in this case equivalent to the
HE) part, and $n(n-1)^2(n-2)/12$ for the HH part. Similarly, the Weyl
tensor has these symmetry properties and is also traceless. This
implies that the $n(n+1)(n+2)(n-3)/12$ independent components of
$C_{[2][2]}$ reorganize into $(n+1)(n-2)/2$ for its EE part,
$(n^2-1)(n-3)/3$ for its EH (or HE) part and $n(n^2-1)(n-4)/12$
for its HH part. Notice that, for $n=4$, the HH part of the Weyl
tensor has no {\it new} independent components, and this is why in 4
dimensions only one electric (corresponding to EE=--HH) and one magnetic
part (EH=HE) arise, compare with subsection \ref{subsec:weyl}.

\section{Super-energy tensors}
\label{sec:s-e}
As an application of the above, for any $t_{\mu_1\dots\mu_m}$ and
any $\vec u$ one can define
\b*
W_t\left(\vec{u}\right)\equiv \frac{1}{2}\left(
[{}^t_{\vec{u}}\underbrace{EE\dots E}_r]^2 +\dots +
[{}^t_{\vec{u}}\underbrace{HH\dots H}_{r}]^2 \right) \geq 0
\e*
where $[\,]^2$ means contraction of all indices (divided by the appropriate
factors\cite{S3}). $W_t\left(\vec{u}\right)$ is obviously a non-negative
quantity and it vanishes iff $t_{\mu_1\dots\mu_m}$ is zero. In fact, in any
ON basis one can write
\b*
W_t(\vec{e}_0)=\frac{1}{2}\,\,\sum_{\mu_1,\dots,\mu_m=0}^{n-1}
|t_{\mu_1\dots\mu_m}|^2 \, .
\e*
Comparing with (\ref{e-me}) we see that $W_t\left(\vec{u}\right)$ has the
{\it mathematical} properties of an energy density; being not an energy
in general, though, it is usually called\cite{Beltesis,Old,S,S3}
the ``super-energy'' (s-e) density of the field $t_{\mu_1\dots\mu_m}$ with
regard to $\vec u$. The natural question arises if one can define an
appropriate s-e analogue of the Maxwell tensor (\ref{emem4}). The answer is
positive and in fact dates back to the gravitational s-e tensors found by
Bel\cite{B1,Beltesis,Old} and independently by Robinson. The universal
generalization of the s-e tensors has recently been achieved \cite{S,S3} as
follows. First of all, the notation $t^{\cal{P}}$ with ${\cal P}=1,\dots,2^r$
is used to denote all duals of $t$, where
${\cal P}=1+\sum_{\U=1}^r 2^{\U -1}\epsilon_{\U}$
and $\epsilon_{\U}$ is one or zero
according to whether the block $[n_{\U}]$ is dualized or not.
One can define a product $\odot$ of an $r$-fold form by itself
resulting in a $2r$-tensor
\b*
(t\odot t)_{\lambda_1\mu_1\dots\lambda_r\mu_r}\equiv
\left(\prod_{\U=1}^{r}\frac{1}{(n_{\U}-1)!}\right)\,
\tilde{t}_{\lambda_1\rho_2\dots\rho_{n_1},\dots ,
\lambda_r\sigma_2\dots\sigma_{n_r}}
\tilde{t}_{\mu_1\hspace{10mm}\dots ,\mu_r}^{\hspace{2mm}\rho_2\dots\rho_{n_1}
,\hspace{5mm}\sigma_2\dots\sigma_{n_r}} \, .
\e*
From each block in $t_{[n_1] \dots [n_r]}$ two indices are obtained
in $(t\odot t )_{\lambda_1\mu_1\dots\lambda_r\mu_r}$. With this at hand,
the general definition of the basic s-e tensor of $t$ is \cite{S,S3}:
\be
T_{\lambda_1\mu_1\dots\lambda_r\mu_r}\left\{t\right\}\equiv
\frac{1}{2}\sum_{{\cal P} =1}^{2^r}
\left(t^{\cal{P}}\odot t^{\cal{P}}\right)_{\lambda_1\mu_1\dots\lambda_r\mu_r}.
\label{set}
\ee
Observe that any dual $t^{\cal P}$ of the original tensor $t(=t^1)$
generates the same basic s-e tensor (\ref{set}).
Therefore, one only needs to consider blocks with
at most $n/2$ indices if $n$ is even, or $(n-1)/2$ if $n$ is odd. 
We also remark that $t_{[n_1] \dots [n_r]}$ could contain $n$-blocks
(with dual 0-blocks) for which the expression (\ref{set}) has no meaning.
However, an $n$-form in $n$ dimensions is trivial and contributes
to the superenergy with a scalar factor given precisely by
its dual 0-form squared. We note that
$T_{\lambda_1\mu_1\dots\lambda_r\mu_r}\left\{t\right\}=
T_{(\lambda_1\mu_1)\dots(\lambda_r\mu_r)}\left\{t\right\}$
and if the tensor $t_{[n_1],\dots,[n_r]}$ is symmetric in the interchange
$[n_{\U}] \leftrightarrow [n_{\U'}]$ ($n_{\U}=n_{\U'}$), then
the s-e tensor (\ref{set}) is symmetric in the interchange of the
corresponding $(\lambda_{\U}\mu_{\U})$- and $(\lambda_{\U'}\mu_{\U'})$-pairs.
Also, if $n$ is even, then (\ref{set}) is traceless in any
$(\lambda_{\U}\mu_{\U})$-pair with $n_{\U}=n/2$. It is remarkable that,
after expanding all duals in (\ref{set}),
one obtains an explicit expression for the s-e tensor which is
{\it independent} of the dimension $n$, see \cite{S3}. Finally, one 
can prove the importnat result that if $t_{[n_1] \dots [n_r]}$ is decomposable 
as in (\ref{decom}), then
\be
T_{\lambda_1\mu_1\dots\lambda_r\mu_r}\left\{t_{[n_1] \dots [n_r]} \right\}=
T_{\lambda_1\mu_1}\left\{\O^{(1)}_{[n_1]} \right\}\dots
T_{\lambda_r\mu_r}\left\{\O^{(r)}_{[n_r]} \right\}.\label{decom2}
\ee

The timelike component of 
$T_{\lambda_1\mu_1\dots\lambda_r\mu_r}\left\{t\right\}$ with respect 
to an observer $\vec u$ is precisely the s-e density 
$W_t\left(\vec{u}\right)$ defined above. More importantly, the s-e 
tensors have the fundamental property that they always satisfy a 
generalization of the dominant energy condition (\ref{dec}) called the 
{\it dominant property}, see section \ref{sec:dp}. The s-e tensor
$T_{\lambda_1\mu_1\dots\lambda_r\mu_r} \{ t\}$ and its derived tensors by
permutation of indices are the {\it only} (up to linear combinations) tensors
quadratic in $t$ and with the dominant property \cite{S3}.
Therefore, $T_{(\lambda_1\mu_1\dots\lambda_r\mu_r)} \{ t\}$ is, up to a
trivial factor, the unique {\it completely symmetric} tensor 
quadratic in $t$ with the dominant property. This property of
general s-e tensors was used
in \cite{BS} to find criteria for the causal propagation of fields on
$n$-dimensional spacetimes.

In 4 dimensions, the s-e tensor of a 2-form $F_{ab}=F_{[ab]}$ is its
Maxwell energy-momentum tensor (\ref{emem4}), and the s-e tensor of an exact
1-form $d\phi$ has the form (in any $n$ after expanding duals)
\be
T_{\mu\nu}\{\nabla_{[1]}\phi \}=\nabla_{\mu}\phi\nabla_{\nu}\phi -
\frac{1}{2}(\nabla_{\rho}\phi\nabla^{\rho}\phi)g_{\mu\nu} \label{ems}
\ee
which is exactly the energy-momentum tensor for a massless
scalar field $\phi$. If we compute the s-e tensor of the Riemann
tensor we get the so-called Bel tensor (see \cite{B4,Beltesis} for $n=4$)
\b*
2\, T_{\alpha}{}^{\beta}{}_{\lambda}{}^{\mu}\left\{R_{[2],[2]}\right\}=
R_{\alpha\rho,\lambda\sigma}
R^{\beta\rho,\mu\sigma}
+\frac{1}{(n-3)!}\,R_{\alpha\rho,
\stackrel{*}{\lambda\sigma_{4}\dots\sigma_n}}
R^{\beta\rho,\stackrel{*}{\mu\sigma_{4}\dots\sigma_n}}
+\\
\frac{1}{(n-3)!}\,
R_{\stackrel{*}{\alpha\rho_{4}\dots\rho_n},\lambda\sigma}
R^{\stackrel{*}{\beta\rho_{4}\dots\rho_n},\mu\sigma}
+\frac{1}{\left[(n-3)!\right]^2}
R_{\stackrel{*}{\alpha\rho_{4}\dots\rho_n},
\stackrel{*}{\lambda\sigma_{4}\dots\sigma_n}}
R^{\stackrel{*}{\beta\rho_{4}\dots\rho_n},
\stackrel{*}{\mu\sigma_{4}\dots\sigma_n}} 
\e*
which, after expanding the duals, becomes independent of $n$:
\bea
B_{\alpha\beta\lambda\mu}\equiv 
T_{\alpha\beta\lambda\mu}\left\{R_{[2],[2]}\right\}=
R_{\alpha\rho,\lambda\sigma}
R_{\beta}{}^{\rho,}{}_{\mu}{}^{\sigma}
+R_{\alpha\rho,\mu\sigma}
R_{\beta}{}^{\rho,}{}_{\lambda}{}^{\sigma}-\nonumber\\
-\frac{1}{2}g_{\alpha\beta}
R_{\rho\tau,\lambda\sigma}R^{\rho\tau,}{}_{\mu}{}^{\sigma}
-\frac{1}{2}g_{\lambda\mu}
R_{\alpha\rho,\sigma\tau}R_{\beta}{}^{\rho,\sigma\tau}+
\frac{1}{8}g_{\alpha\beta}g_{\lambda\mu}
R_{\rho\tau,\sigma\nu}
R^{\rho\tau,\sigma\nu} .\label{Bel}
\eea
The s-e tensor of the Weyl curvature has exactly the same expression 
as (\ref{Bel}) by replacing $C_{\alpha\rho,\lambda\sigma}$ for
$R_{\alpha\rho,\lambda\sigma}$, and is called the generalized
Bel-Robinson tensor \cite{B1,Old,Beltesis,S3}. In 4 and 5 dimensions it is 
completely symmetric, and also traceless for $n=4$, but not in general. 
Of course, the Bel and Bel-Robinson tensors coincide in Ricci-flat 
spacetimes. The dominant property of
the Bel-Robinson tensor was used by Christodoulou and Klainerman \cite{CK}
in their study of the global stability of Minkowski spacetime, and in
\cite{BoS} to prove the causal propagation of gravity in vacuum for 
$n=4$.

By using the second Bianchi identity $\nabla_{[\nu}R_{\alpha\beta]\lambda\mu}=0$,
it follows from (\ref{Bel})
\be
\nabla_{\alpha}B^{\alpha\beta\lambda\mu}=
R^{\beta\hspace{1mm}\lambda}_{\hspace{1mm}\rho\hspace{2mm}\sigma}
j^{\mu\sigma\rho}+R^{\beta\hspace{1mm}\mu}_{\hspace{1mm}\rho\hspace{2mm}\sigma}
j^{\lambda\sigma\rho}-\frac{1}{2}g^{\lambda\mu}
R^{\beta}_{\hspace{1mm}\rho\sigma\gamma}j^{\sigma\gamma\rho}\label{divbel}
\ee
where $j_{\lambda\mu\beta}\equiv
\nabla_{\lambda}R_{\mu\beta}-\nabla_{\mu}R_{\lambda\beta}$. This is 
analogous to the formula shown for $\tau_{\mu\nu}$ in subsection 
\ref{subsec:e-m} and, again, it leads to the interesting result
\b*
j_{\lambda\mu\beta}=0 \hspace{5mm} \Longrightarrow \hspace{5mm}
\nabla_{\rho}B^{\rho\beta\lambda\mu}=0.
\e*
In particular, the Bel (and Bel-Robinson) tensor is divergence-free in
Ricci-flat or Einstein spacetimes. If there is a Killing vector $\vec 
\xi$ we get 
\be
j_{\lambda\mu\beta}=0 \hspace{5mm} \Longrightarrow \hspace{5mm}
\nabla_{\rho}J^{\rho}=0, \hspace{1cm} J^{\rho}(R_{[2][2]};\xi)\equiv
B^{\rho\beta\lambda\mu}\xi_{\beta}\xi_{\lambda}\xi_{\mu} \label{J}
\ee
so that again the {\it divergence-free current} 
$\vec J$ can be used to construct conserved quantities with the Gauss
theorem. When the charge current $\vec j$ for the electromagnetic 
field, or the $j_{\lambda\mu\beta}$ for the gravitational field, are 
not zero the corresponding currents are no longer divergence-free. In 
the case of classical Electromagnetism this was interpreted as a 
signal of the existence of energy-momentum associated to the charges 
creating the field and appearing in $\vec j$, see e.g.\cite{LL}. When 
one constructs the energy-momentum tensor corresponding to these 
charges and uses the field equations, the {\it total} energy-momentum 
tensor is again divergence-free\cite{LL} and can be used to construct the 
conserved {\it mixed} quantities. Can one find similar results for 
the case of the gravitational field and the Bel tensor?

A positive answer to this question requires, at least, the definition 
of s-e tensors of the Bel type for physical fields. However, this can 
be achieved by using our general definition (\ref{set}) and following 
an idea launched long ago by Chevreton \cite{Ch} and recently retaken 
by other authors \cite{B5,Tey} in the case of Minkowski spacetime. 
The combined idea was put forward in \cite{S,S2,S3} and simply uses 
the covariant derivatives of the physical fields to construct the 
appropriate s-e tensors. As an outstanding example, we can define
the basic s-e tensor of the electromagnetic field as that 
corresponding to the double (1,2)-form $\nabla_{\alpha}F_{\mu\nu}$
\b*
E_{\alpha\beta\lambda\mu}\equiv
T_{\alpha\beta\lambda\mu}\{\nabla_{[1]}F_{[2]}\}=
\nabla_{\alpha}F_{\lambda\rho}\nabla_{\beta}F_{\mu}{}^{\rho}+
\nabla_{\alpha}F_{\mu\rho}\nabla_{\beta}F_{\lambda}{}^{\rho}-\nonumber \\
g_{\alpha\beta}\nabla_{\sigma}F_{\lambda\rho}
\nabla^{\sigma}F_{\mu}{}^{\rho} -\frac{1}{2}g_{\lambda\mu}
\nabla_{\alpha}F_{\sigma\rho}\nabla_{\beta}F^{\sigma\rho}+
\frac{1}{4}g_{\alpha\beta}g_{\lambda\mu}
\nabla_{\tau}F_{\sigma\rho}\nabla^{\tau}F^{\sigma\rho} 
\e*
whose symmetry properties are
$E_{\alpha\beta\lambda\mu}=E_{(\alpha\beta)(\lambda\mu)}$ and has zero 
divergence in flat spacetimes \cite{S3}. The tensor
$E_{(\alpha\beta\lambda\mu)}$ is unique in the sense commented on 
above and coincides with the symmetric part of Chevreton's \cite{Ch}, 
and can be used to construct conserved quantities along shock-waves 
with continuous $F_{[2]}$, see \cite{L,S,S3}. This is a first 
indication of the possibility of the interchange of `superenergy' 
quantities between different fields. However, the complete analysis 
in the Einstein-Maxwell case is still to be done.

Another exact and interesting result was achieved, though, in the simpler 
case of a scalar field $\phi$ coupled to gravity. In order to get the 
basic s-e tensor of a {\it massless} scalar field one can use the double symmetric
(1,1)-form $\nabla_{\alpha}\nabla_{\beta}\phi$ as starting object, so that
the corresponding tensor (\ref{set}) becomes 
\bea
S_{\alpha\beta\lambda\mu}\equiv
T_{\alpha\beta\lambda\mu}\{\nabla_{[1]}\nabla_{[1]}\phi\}=
\nabla_{\alpha}\nabla_{\lambda}\phi \nabla_{\mu}\nabla_{\beta}\phi 
+\nabla_{\alpha}\nabla_{\mu}\phi \nabla_{\lambda}\nabla_{\beta}\phi - 
\hspace{1cm} \nonumber \\
-g_{\alpha\beta}\nabla_{\lambda}\nabla^{\rho}\phi\nabla_{\mu}\nabla_{\rho}\phi 
- g_{\lambda\mu}\nabla_{\alpha}\nabla^{\rho}\phi\nabla_{\beta}\nabla_{\rho}\phi 
+\frac{1}{2} g_{\alpha\beta}g_{\lambda\mu}
\nabla_{\sigma}\nabla_{\rho}\phi\nabla^{\sigma}\nabla^{\rho}\phi \, .
\label{ses}
\eea
In fact this tensor was previously found by Bel \cite{B5} and Teyssandier
\cite{Tey} in Special Relativity ($n=4$). Its
symmetry properties are
$S_{\alpha\beta\lambda\mu}=S_{(\alpha\beta)(\lambda\mu)}=
S_{\lambda\mu\alpha\beta}$. Concerning its divergence, a straightforward
computation using the Ricci identity\cite{HE} leads to
\b*
\nabla_{\alpha}S^{\alpha}{}_{\beta\lambda\mu}=
2\nabla_{\beta}\nabla_{(\lambda}\phi
R_{\mu)\rho}\nabla^{\rho}\phi -g_{\lambda\mu} R^{\sigma\rho}
\nabla_{\beta}\nabla_{\rho}\phi\nabla_{\sigma}\phi -\nonumber\\
-\nabla_{\sigma}\phi \left(2\nabla^{\rho}\nabla_{(\lambda}\phi \,
R^{\sigma}_{\mu)\rho\beta} +g_{\lambda\mu}
R^{\sigma}_{\rho\beta\tau}\nabla^{\rho}\nabla^{\tau}\phi\right)
\e*
so that all the terms on the righthand side involve
components of the Riemann tensor. Thus, in any flat region of the
spacetime (that is, with vanishing Riemann tensor, so that there is no 
gravitational field), the s-e tensor of the massless scalar field (\ref{ses})
is divergence-free. This leads to conserved currents of type (\ref{J})
for the scalar field in flat spacetimes\cite{S2,S3,Tey}. For a deeper
study of these in Special Relativity see\cite{Tey}.

The situation is therefore that the Bel tensor is divergence-free
in Ricci-flat spacetimes, and the s-e tensor (\ref{ses}) is divergence-free in
the absence of gravitational field. These divergence-free properties lead to
conserved currents of type (\ref{J}) whenever there are symmetries in the
spacetime, see \cite{S3} for a lengthy discussion. Thus, the important question
arises if one can combine the Bel tensor with the s-e tensor 
(\ref{ses}) to produce a conserved
current in the {\it mixed} case when there are both a scalar field and the
curvature that it generates. Assume therefore that the
Einstein-Klein-Gordon equations for a minimally coupled massless scalar
field hold \footnote{Actually, the result holds true for massive fields 
too.\cite{S2,S3}}, and that there is a Killing vector $\vec \xi$ in the
spacetime. A simple calculation\cite{S2,S3} leads to
\b*
\nabla_{\alpha}J^{\alpha}=0 , \hspace{3mm} \mbox{where} \hspace{3mm}
J^{\alpha}\left(R_{[2][2]},\nabla_{[1]}\nabla_{[1]}\phi ; \xi\right)\equiv
\left(B^{\alpha\beta\lambda\mu}+
S^{\alpha\beta\lambda\mu}\right)\xi_{\beta}\xi_{\lambda}\xi_{\mu}\, .
\e*
Notice that only the completely symmetric parts of $B$ and $S$ are
relevant here. This provides conserved quantities proving the exchange
of s-e properties between the gravitational and scalar fields, because neither
$B^{\alpha\beta\lambda\mu}\xi_{\beta}\xi_{\lambda}\xi_{\mu}$ nor
$S^{\alpha\beta\lambda\mu}\xi_{\beta}\xi_{\lambda}\xi_{\mu}$ are
divergence-free separately (if the other field is present!).

\section{Causal tensors}
\label{sec:dp}
As mentioned in Section \ref{sec:s-e}, the s-e tensors satisfy a 
generalization of the dominant energy condition (\ref{dec})
given in the following
\begin{defi}
A tensor $T_{\mu_{1}...\mu_{s}}$ is said to
have the \underline{dominant property} if
$$T_{\mu_{1}...\mu_{s}}v_{1}^{\mu_{1}}...v_{s}^{\mu_{s}}\geq 0$$ for any set
$\vec{v}_{1}$,..., $\vec{v}_{s}$ of causal future-pointing vectors.
The set of tensors with the dominant property is denoted by $\DP$, and
$-\DP$ will mean the set of tensors $T_{\mu_{1}...\mu_{s}}$ such that
$-T_{\mu_{1}...\mu_{s}}\in \DP$.
\label{DP}
\end{defi}
Notice that $\DP \cap -\DP =\{0\}$.
The non-negative real numbers are considered a special case
with the dominant property, $\R^+ \subset \DP$.
Rank-1 tensors with the dominant property are the past-pointing
causal vectors, while those in $-\DP$ are the future-directed ones. For
rank-2 tensors, the dominant property is exactly (\ref{dec}),
usually called the dominant energy
{\it condition} for it is a requirement placed on physically
acceptable energy-momentum tensors \cite{Ple,HE}. 
As in the case of past- and future-pointing vectors, any statement
concerning $\DP$ has its counterpart concerning $-\DP$, and they will be
omitted sometimes. The elements of $\DP\cup -\DP$ will thus be called
``causal tensors'', and this set has the mathematical structure of 
a {\it cone} because
\b*
T_{\mu_{1}...\mu_{s}},\, S_{\mu_{1}...\mu_{s}}\in \DP \hspace{4mm}
\Longrightarrow \hspace{4mm}
a\, T_{\mu_{1}...\mu_{s}}+b\, S_{\mu_{1}...\mu_{s}}\in \DP \hspace{2mm}
\forall a,b\in \R^+\, .
\e*
Therefore, this seems to be an adequate generalization of the solid 
Lorentz cone to tensors of arbitrary rank. A natural question that 
arises is: what are the boundaries of these solid cones, or in other 
words, which are the higher-rank generalizations of the null cone?
Actually, we also have $\forall T_{1},T_{2}\in \DP$, $T_{1}\otimes 
T_{2}\in \DP$ \cite{BerS}, so that $\DP\cup -\DP$ is a sort of graded 
algebra of solid cones.

The main properties of $\DP$ are summarized in what follows, see 
also\cite{BerS} and Bergqvist's contribution to this volume. To that 
end, the following notations will be used: a) for any two tensors
$T_{\mu_{1}...\mu_{r}}^{(1)}$ and $T_{\mu_{1}...\mu_{s}}^{(2)}$, we 
write
$$
(T^{(1)}\, {}_i\!\times_j\, T^{(2)})_{\mu_1\dots\dots \mu_{r+s-2}}\equiv
T^{(1)}_{\mu_{1}...\mu_{i-1}\rho\mu_{i}...\mu_{r-1}}
T^{(2)}_{\mu_{r}...\mu_{r+j-2}}{}^{\rho}{}_{\mu_{r+j-1}...\mu_{r+s-2}}
$$
so that a contraction with the $i^{th}$ index of the first tensor and
the $j^{th}$ of the second is taken. b) In particular, for any two 
rank-2 tensors $A_{\mu\nu}$ and $B_{\mu\nu}$ we will simply write 
$A\times B\equiv A\, {}_1\!\times_1 B$, which is another rank-2 
tensor. c) For any two tensors $T_{\mu_{1}...\mu_{s}},\, 
S_{\mu_{1}...\mu_{s}}$ of the same rank we put $T\cdot S \equiv
T_{\mu_{1}...\mu_{s}} S^{\mu_{1}...\mu_{s}}$.

Several characterizations of the set $\DP$ are the following
\b*
T_{\mu_{1}...\mu_{s}}\in \DP \hspace{2mm} \Longleftrightarrow \hspace{2mm}
T_{\mu_{1}...\mu_{s}}u_{1}^{\mu_{1}}...u_{s}^{\mu_{s}}> 0, \,\,\,
\forall \hspace{1mm}\mbox{\underline{timelike}}\hspace{1mm}
\vec{u}_{1},\dots , \vec{u}_{s} 
\hspace{3mm} \Longleftrightarrow \\ 
\hspace{2mm} \Longleftrightarrow \hspace{2mm}
T_{\mu_{1}...\mu_{s}}k_{1}^{\mu_{1}}...k_{s}^{\mu_{s}}\geq 0, \,\,\,
\forall \hspace{1mm}\mbox{\underline{null}}\hspace{1mm}
\vec{k}_{1},\dots , \vec{k}_{s} 
\hspace{2mm} \Longleftrightarrow \hspace{2mm}\\
\hspace{2mm} \Longleftrightarrow \hspace{2mm}
T {}_i\!\times_j S \in \DP, \,\,\, \forall S\in -\DP \hspace{3mm}
\Longleftrightarrow \hspace{2mm}
T {}_i\!\times_j t \in -\DP, \,\,\, \forall t\in \DP \\
\Longleftrightarrow \hspace{2mm}
T_{0...0}\geq \left| T_{\alpha _{1}...\alpha_{s}}\right| \,\, 
\mbox{in any ON basis} \,\,\,
\left\{\vec{e}_{\mu}\right\} \,\,\, \mbox{with
a future} \,\,\, \vec{e}_{0},
\e*
while some characterizations of causal tensors are
\b*
0\neq T\in \DP\cup -\DP \hspace{2mm} \Longleftrightarrow \hspace{2mm}
0\neq T {}_i\!\times_i T \in -\DP \hspace{3mm} \mbox{for some} \, i \\
\Longleftrightarrow \hspace{3mm}
0\neq T {}_i\!\times_j T \in -\DP \,\, \forall i,j \hspace{1cm}
\e*
or the extreme cases given by
\b*
T \, {}_i\!\times_i T =0 \hspace{3mm} \forall \, i \hspace{4mm}
\Longleftrightarrow \hspace{3cm}\\
\Longleftrightarrow \hspace{2mm}
T=\bm{k}_{1}\otimes \dots \otimes \bm{k}_{s} \,\,
\mbox{for a set of \underline{null}}\hspace{1mm}
\bm{k}_{1},\dots , \bm{k}_{s} \Longrightarrow T\in \DP\cup -\DP .
\e*
If $T_{\mu_{1}...\mu_{s}}$ is antisymmetric in any 
two indices then $T_{\mu_{1}...\mu_{s}}\notin \DP\cup -\DP$. In 
particular, $p$-forms $\bm{\S}$ with $p>1$ cannot be in $\DP\cup -\DP$. 
Nevertheless, we will still use the name {\it causal} $p$-form for the 
single forms $\bm{\S}$ such that $\S \cdot \S \leq 0$.
Recall that a $p$-form $\Omega_{\nu_{1}...\nu_{p}}=\Omega_{[\nu_{1}...\nu_{p}]}$
is called \underline{simple} if it is a product of $p$ linearly independent
1-forms $\bm{\omega}^{1},\dots ,\bm{\omega}^{p}$, i.e.
$\bm{\Omega}=\bm{\omega}^{1}\wedge ...\wedge \bm{\omega}^{p}$.
A $p$-form $\Omega_{[p]}$ is simple iff
$\Omega_{\stackrel{*}{[N-p]}}$ is simple, and if and only if
$\Omega_{[p]}\, {}_1\!\!\times_1 \Omega_{\stackrel{*}{[N-p]}}=0$,
see e.g. \cite{Sc,PR}. According to the definition given in 
subsection \ref{subsec:singleEH}, a $p$-form $\bm{\Omega}$ is null if it is simple 
and causal with $\Omega \cdot \Omega = 0$.

Let us introduce the following classes: $\SE$ will denote the set of all 
s-e tensors (\ref{set}) introduced in section \ref{sec:s-e} plus the 
rank-2 tensors of type $fg_{\mu\nu}$ with $f\leq 0$; by $\SS$ is meant 
the set of s-e tensors of {\it simple} $p$-forms plus the tensors of type
$fg_{\mu\nu}$ with $f\leq 0$; finally $\NS$ denotes the set of s-e tensors
of {\it null} $p$-forms. By $-\SE$, $-\SS$ and $-\NS$ we mean the sets 
of tensors $T$ such that $-T\in \SE$, $\SS$ or $\NS$, respectively.
Notice that all tensors in $\NS$ and $\SS$ are of 
rank 2, and that $\NS \subset\SS\subset\SE$. Furthermore, $\SE\subset \DP$,
as was proved in \cite{Ber} for $n=4$ using spinors and later for 
general $n$ in\cite{S3,PP}. An important result of relevance in General 
relativity is\cite{BerS}
\begin{prop}
In dimension $n\leq 4$, $T_{\mu\nu}\in \SE \Longrightarrow
(T\times T)_{\mu\nu}=h^{2}g_{\mu\nu}$.
\label{prop:TxT}
\end{prop}
This result does {\it not} hold for $n>4$, and one wonders what is its 
adequate generalization to any $n$. Similarly, it is desirable to know 
the converse of Proposition \ref{prop:TxT}, as this is the basis of 
the famous Rainich conditions\cite{R,MW,ex} which allow to prove the 
existence of an electromagnetic field based only on the underlying 
geometry if $T_{\mu}{}^{\mu}=0$, see (\ref{raiem}).
These questions are to be answered now and in the next section, and 
are actually related to that asked before about the `higher-rank' 
generalizations of the null cone.

To begin with, we have\cite{BerS}
\begin{prop}
$T_{\mu\nu}\in \SS \Longrightarrow (T\times T)_{\mu\nu}=h^{2}g_{\mu\nu}$.
\label{prop:txt}
\end{prop}
This result holds in arbitrary $n$ and generalizes Proposition 
\ref{prop:TxT}. Furthermore, we also have the following fundamental 
result
\begin{theo}
In $n$ dimensions, any symmetric rank-2 tensor $S_{(\mu\nu)}=S_{\mu\nu}\in \DP$
can be written
\be
S_{\mu\nu}=T_{\mu\nu}\{\Omega_{[1]}\}+...+T_{\mu\nu}\{\Omega_{[n-1]}\}-
\alpha g_{\mu\nu}
\label{bb2}
\ee
where $\alpha \geq 0$ and $T_{\mu\nu}\{\Omega _{[p]}\}\in \SS$
are the superenergy tensors of \underline{simple} and \underline{causal}
$p$-forms $\Omega_{[p]}$, $p=1,...,n-1$
such that for $p=2,\dots ,n-1$ they have
the structure $\Omega_{[p]}=\bm{k}^{1}\wedge \dots \wedge \bm{k}^{p}$
where $\bm{k}^{1},\dots ,\bm{k}^{p}$ are appropriate {\em null} 
1-forms. The number of tensors on the righthand side of (\ref{bb2}) 
and the structure of the $\Omega_{[p]}$
depend on the particular $S_{\mu\nu}$ as follows: if $S_{\mu\nu}$ has
$n-q\geq 1$ null eigenvectors $\vec{k}^1,\dots,\vec{k}^{n-q}$ then {\em at least}
$T_{\mu\nu}\{\Omega_{[n-q]}\}$, with
$\Omega_{[n-q]}=\bm{k}^{1}\wedge \dots \wedge \bm{k}^{n-q}$,
must appear in (\ref{bb2}), and possibly those to its right in (\ref{bb2}).
If it has no null eigenvectors, then {\em at least}
$T_{\mu\nu}\{\Omega_{[1]}\}$ appears in (\ref{bb2}), and possibly those to its
right, and $\Omega_{[1]}$ is timelike. 
\label{theo:bb2}
\end{theo}
{\it Remark:} As already mentioned, the s-e tensor of the dual of a
$p$-form is identical with that of the $p$-form itself. Thus, in the sum
(\ref{bb2}) there are two s-e tensors of 1-forms, namely
$T_{\mu\nu}\{\Omega_{[1]}\}$ and $T_{\mu\nu}\{\Omega_{\stackrel{*}{[1]}}\}
=T_{\mu\nu}\{\Omega_{[n-1]}\}$, but the first one corresponds to
a {\em causal} 1-form and the second to a {\em spacelike} one.
Similar remarks apply to the 2-forms $\Omega_{[2]}$
and $\Omega_{\stackrel{*}{[n-2]}}$, and so on. 

Theorem \ref{theo:bb2} means that the elements in
$\SS$ can be used to build up $\DP$ and, in this sense, $\SS$ seems a 
generalization of the null cone to rank-2 tensors. One can compare 
Theorem \ref{theo:bb2} with the standard result that any causal and 
past-directed vector $\vec v$ can be expressed as the combination of 
two past-pointing null vectors $\vec{k}^1,\vec{k}^2$: $\vec v=\vec{k}^1+\alpha 
\vec{k}^2$ with $\alpha \geq 0$. Recall that this can be achieved in
infinitely many ways. In the same manner, a symmetric $S_{\mu\nu}\in \DP$
can be expressed as a sum of $n$ elements of $\SS$ in many ways.
In the above Theorem \ref{theo:bb2}, however, the representation of
$S_{\mu\nu}\in \DP$ has been chosen in a canonical way
in relation with the null eigenvectors of $S_{\mu\nu}$.

Theorem \ref{theo:bb2} allows also to prove the converse of 
Proposition \ref{prop:txt}, see\cite{BerS}, in the following sense
\begin{theo}
If $T_{\mu\nu}$ is symmetric and
$(T\times T)_{\mu\nu}=fg_{\mu\nu}$
then:

{\em (a)} $f=0 \Longrightarrow T_{\mu\nu}\in \NS\cup -\NS$ and $T_{\mu\nu}=
k_{\mu}k_{\nu}$ for a null $\bm{k}$.

{\em (b)} $f\neq 0 \Longrightarrow T_{\mu\nu}\in \SS\cup -\SS$. Moreover,
$\epsilon T_{\mu}{}^{\mu}/\sqrt{T_{\rho\sigma}T^{\rho\sigma}/n}$ 
takes one of the values $\left\{n,(n-2),\dots,(2p-n),\dots,(2-n)\right\}$
according to the rank $p$ of the simple $p$-form $\Omega_{[p]}$ generating
$T_{\mu\nu}\in \SS\cup -\SS$, where $\Omega_{[p]}$ is causal
of the type used in Theorem \ref{theo:bb2} and $\epsilon^2=1$.
\label{theo:fund}
\end{theo}
Proposition \ref{prop:TxT} and Theorem \ref{theo:fund} can be 
combined to give
\begin{coro}
If $n\leq 4$ then $\SE=\SS$. 
\end{coro}
For $n=4$ this means that the
energy-momentum tensor (\ref{emem4}) of any Maxwell field $F_{[2]}$, be 
it simple or not,
coincides with the energy-momentum of (possibly) another \underline{simple}
2-form. This is well known, related to the  duality rotations 
\cite{PR,PlPr}, and forms the basis of the algebraic Rainich 
conditions\cite{R,PR,MW,ex}, statable as
\begin{coro}[Algebraic classical Rainich's conditions]
In $n=4$, a tensor $T_{\mu\nu}$ is (up to sign) {\em algebraically}
the energy-momentum tensor of a 2-form, that is, it takes the form 
(\ref{emem4}),
if and only if $(T\times T)_{\mu\nu}=fg_{\mu\nu}$, 
$T_{\mu\nu}=T_{(\mu\nu)}$ and 
$T_{\mu}{}^{\mu}=0$,
which are relations (\ref{raiem}).
\end{coro}

All the above can also be seen from the point of view of Lorentz 
transformations and some of its generalizations. To that end, let us 
introduce the following terminology:
$T_{\mu}{}^{\nu}$ is said to be a {\it null-cone preserving map}
if $k^{\mu} T_{\mu}{}^{\nu}$ is null for any null vector $\vec{k}$. A map that
preserves the null cone is said to be {\it orthochronus} (respectively
{\it time reversal}) if it keeps (resp. reverses) the cone's
time orientation, and is called {\it proper}, {\it improper} or
{\it singular} if $\det(T_{\mu}{}^{\nu})$ is positive, negative, or zero,
respectively. If the map is proper and orthochronus then it is called
{\it restricted}. A null-cone preserving map is {\it involutory}
if $T_{\mu}{}^{\nu}=(T^{-1}){}_{\mu}{}^{\nu}$, and {\it bi-preserving} if
$T_{\mu}{}^{\nu} k_{\nu}$ is also null for any null 1-form $\bm{k}$.
Notice that involutory null-cone
preserving maps are necessarily non-singular.
The following lemma gives a geometrical interpretation
to some previous results.
\begin{lem}
$(T {}_2\!\times_2 T)_{\mu\nu}=fg_{\mu\nu} \Longleftrightarrow T_{\mu}{}^{\nu}$
is a null cone preserving map.
\label{lem:TxT=fg}
\end{lem}
This allows to prove\cite{BerS} that, in fact, $(\SS\cup -\SS) \setminus 
(\NS\cup -\NS)$ is the set of 
tensors proportional to involutory Lorentz transformations, while 
$\NS\cup -\NS$ is a subset of the {\it singular} null-cone bi-preserving 
maps. They can be easily classified into 
orthochronus or time-reversal, and proper or improper using the above 
results, see \cite{BerS}.

\section{Generalization of algebraic Rainich conditions}
\label{sec:rainich}
Finally, let us see how can one use all the previous results to 
provide algebraic Rainich-like conditions in any dimension $n$.
To start with, for a massless scalar field we have
\begin{coro}
In $n$ dimensions, a tensor $T_{\mu\nu}$ is {\em algebraically}
the energy-momentum tensor (\ref{ems}) of a minimally coupled massless scalar
field $\phi$ if and only if $T_{\mu\nu}\in \SS$ and $T_{\mu}{}^{\mu} =\beta
\sqrt{T_{\mu\nu}T^{\mu\nu}/n}$
where $\beta =\pm (n-2)$. Moreover, $d\phi$ is spacelike or timelike
if $T_{\mu}{}^{\mu}\ne 0$ and $\beta =n-2$ or $\beta =2-n$, respectively,
and null if $T_{\mu}{}^{\mu} =0=T_{\mu\nu}T^{\mu\nu}$.
\end{coro}
This simple, complete and general result is to be compared with 
alternative approaches to the problem \cite{Per,Ku,Pen1,Pen2} for
$n=4$. Of course, the Corollary follows almost immediately from 
Theorem \ref{theo:fund}.

We can also attack the problem of algebraic Rainich conditions for a
perfect fluid in arbitrary dimension $n$.
\begin{coro}
A tensor $T_{\mu\nu}$ is {\em algebraically}
the energy-momwntum tensor of a perfect fluid satisfying the dominant
energy condition if and only if
\be
T_{\mu\nu}=-\frac{\lambda}{2}g_{\mu\nu}+\mu T_{\mu\nu}\{v_{[1]}\} \label{pf}
\ee
where $\lambda ,\mu \geq 0$ and $\bm{v}$ is timelike, and therefore 
the tensor $T_{\mu\nu}\{v_{[1]}\}\in \SS$
is intrinsically characterized according to its trace,
see Theorem \ref{theo:fund}. The velocity vector of the fluid, its energy
density and its pressure are given by $\vec u=-\vec v/(v\cdot v)$,
$\rho =(-\mu (v\cdot v)+\lambda)/2$ and $P=-(\mu (v\cdot v)+\lambda)/2$,
respectively.
\end{coro}
The proof is quite simple again. Recall that a perfect fluid has the Segre
type $\{1,(1\dots 1)\}$, so that
\be
T_{\mu\nu}=(\rho +P)u_{\mu} u_{\nu} +Pg_{\mu\nu} \label{pf2}
\ee
where $(u\cdot u)=-1$. Thus, if (\ref{pf}) holds it is obvious that 
$T_{\mu\nu}$
takes the form (\ref{pf2}). Conversely, if (\ref{pf2}) holds, then
$T_{\mu\nu}-T_{\mu\nu}\{u_{[1]}\}$ has every null $\vec k$ as eigenvector,
as can be trivially checked. Therefore, $T_{\mu\nu}-T_{\mu\nu}\{u_{[1]}\}$
is proportional to the metric, as follows from Theorem \ref{theo:bb2} and the
proportionality factor is obtained from the trace.
In fact, we can rederive the conditions found for $n=4$ in\cite{CF,M}
generalized to $n$ dimensions as follows. From (\ref{pf}) we get
$$
(T\times T)_{\mu\nu}=
\lambda T_{\mu\nu}-\rho Pg_{\mu\nu}
$$
and also $n(T\times T)_{\mu}{}^{\mu}-(T_{\mu}{}^{\mu})^2\geq 0$,
$T_{\mu}{}^{\mu}\geq n\lambda/2$
and $T_{\mu\nu}w^{\mu}w^{\nu}\geq \lambda /2$ for all timelike $\vec w$.

As a final example, let us consider the case of dust ($P=0$ perfect 
fluids). Of course, the characterization of dust can be deduced from 
the previous one by setting $P=0$. However, there are some particular 
cases in which some stronger results can be derived. For instance
\begin{coro}
In 5 dimensions, a tensor $T_{\mu\nu}$ is {\em algebraically}
the energy-momwntum tensor of a dust $T_{\mu\nu}=\rho u_{\mu} u_{\nu}$
where $(u\cdot u)=-1$ and $\rho\geq 0$ if and only if $T_{\mu\nu}$ is 
the s-e tensor of a 2-form $F_{[2]}$ with no null eigenvector.
\end{coro}

Many other results of this type can be obtained.

\section*{Acknowledgments}
I am grateful to the organizers for inviting me to the ERE-00 
meeting. Some parts of this paper have been obtained as result of a 
collaboration with Dr. G\"oran Bergqvist. I thank Dr. Ra\"ul Vera for correcting
some errors and reading the manuscript. Financial support from the
Basque Country University under project
number UPV172.310-G02/99 is acknowledged.

\end{document}